\documentstyle[preprint,aps,prc,floats,psfig,tighten,epsfig,axodraw,epic]{revtex}
\draft
\renewcommand{\thefootnote}{\arabic{footnote}}

\def\epem{e^+e^-}

\def\lsim{\mathrel{\raise.3ex\hbox{$<$\kern-.75em\lower1ex\hbox{$\sim$}}}}
\def\gsim{\mathrel{\raise.3ex\hbox{$>$\kern-.75em\lower1ex\hbox{$\sim$}}}}

\let\jnfont=\rm
\def\NPB#1,{{\jnfont Nucl.\ Phys.\ B }{\bf #1},}
\def\PLB#1,{{\jnfont Phys.\ Lett.\ B }{\bf #1},}
\def\PRD#1,{{\jnfont Phys.\ Rev.\ D }{\bf #1},}
\def\PRL#1,{{\jnfont Phys.\ Rev.\ Lett.\ }{\bf #1},}

\begin{document}

\draft
\renewcommand{\thefootnote}{\arabic{footnote}}

\preprint{
\vbox{\hbox{\bf MADPH-01-1230}
      \hbox{\bf WM-01-107}
      \hbox{\bf hep-ph/0106277}
      \hbox{June, 2001}}}

\title{\bf Search for $t \to ch$ at $\epem$ Linear Colliders}
\author{Tao Han\footnote{than@pheno.physics.wisc.edu}
and Jing Jiang\footnote{jiang@pheno.physics.wisc.edu}}
\address{Department of Physics, University of Wisconsin,\\ 
1150 University Avenue, Madison, WI 53706, USA}
\author{Marc Sher\footnote{sher@physics.wm.edu}}
\address{Nuclear and Particle Theory Group, Physics Department, \\
College of William and Mary, Williamsburg, VA 23187, USA}
\maketitle

%\vskip 0.5cm

\begin{abstract}
We study the rare top-quark decay $t\rightarrow ch$, 
where $h$ is a generic Higgs boson, at a linear collider.  
If kinematically accessible, 
all models contain this decay at some level due to
quark flavor mixing. Some models, such as Model III of the
two-Higgs doublet model, have a tree-level 
top-charm-Higgs coupling, and the branching ratio is close to $0.5\%$.  
Others, such as the MSSM, have a coupling induced at one-loop, 
and can have a branching ratio in the range of 
$10^{-5}-5\times 10^{-4}$.   
We find that a linear collider of $\sqrt{s} = 500$ GeV and 
a luminosity of $500$ fb$^{-1}$ will begin to be sensitive to 
this range of the coupling.

\end{abstract}
\newpage

\section{Introduction}

Two of the most important unanswered puzzles of elementary
particle physics are the nature of electroweak symmetry breaking and the 
origin of flavor.  Both are implemented in the
Standard Model of particle physics in an {\it ad hoc} manner.  
Often, extensions of the Standard Model (SM) 
treat these two problems separately; 
extended Higgs models do not address the
flavor problem and flavor symmetry models do not substantively 
address the nature of electroweak symmetry breaking.  
Yet, there are no dynamical models satisfactorily addressing
the two puzzles coherently.
Since the top quark is so heavy, with a mass on the
order of the electroweak symmetry breaking scale, it is
tempting to consider the possible special role of the top 
quark in the electroweak symmetry breaking sector and in
the quark flavor sector, and its interactions could shed 
light on both problems. 
Flavor-changing-neutral-current (FCNC) decays of the top 
quark \cite{eilam} could be very sensitive to various 
extensions of the Standard Model that address both of these puzzles.  

Most studies of flavor-changing-neutral-current top-quark decays 
have examined the decay of the top quark into a charmed quark 
and a gauge boson \cite{tcv} (either a gluon, photon or a $Z$).  
However, the flavor-changing-neutral-current decay most likely 
to shed light on the nature of electroweak symmetry breaking 
is the decay of the top quark into a charmed quark
and a Higgs boson \cite{eilam,mele} (or the decay of the Higgs 
boson into a top and a charmed quark if kinematically preferred). 
It is this interaction that we study in this paper.

All models will contain a top-charm-Higgs (TCH) vertex at some
level\footnote{Whether a TCH vertex exists does, in part, depend on
nomenclature. Strictly speaking, 
one should consider the ``Higgs" to be the other member of the
isodoublet containing the longitudinal components of the weak vector
bosons; after all, 
it is this field that is associated with spontaneous symmetry
breaking.  With
that definition, the Yukawa couplings are necessarily flavor-diagonal.
Here, we use
the word ``Higgs" to indicate a scalar field whose mass eigenstates
contain some part
of the field associated with spontaneous symmetry breaking.}.  
Even in the Standard
Model, a one-loop diagram gives a contribution, but it is GIM-suppressed
and extremely small \cite{mele}.  
In extensions of the Standard Model, however,
such a vertex
can occur either at tree level or at a non-GIM-suppressed one-loop level.
        
One of the simplest extensions of the Standard Model has a tree-level TCH
vertex.  This is the general two-Higgs doublet model, referred to as Model
III, in
which no discrete symmetries are imposed to avoid flavor-changing neutral
currents.  In this case, the quark Yukawa coupling  matrices consist of
two parts,
$y=y_1+y_2$, where $y_i$ is the coupling to the i'th Higgs doublet.  Since
diagonalizing $y$ will not, in general, diagonalize $y_1$ and $y_2$, there
will be
tree-level flavor-changing neutral currents.  Cheng and
Sher \cite{chengsher} noted
that if one wishes to avoid fine-tuning, then the observed structure of
the mass
and mixing hierarchy suggests that the FCNC vertex $\xi_{ij}\ \bar q_iq_j h$
is given by the geometric mean of the Yukawa couplings of 
the quark fields\footnote{We note that some authors may have
used a different normalization for $\lambda_{ij}$ by a factor of
$\sqrt{2}$ larger than ours. Since the ansatz is order of magnitude
only, either definition is acceptable.  Similarly, we have not 
worried about the running mass effects at different renormalization
scales, and we simply take the pole masses for $m_t$ and $m_c$ for
our analysis.}
\begin{equation}
 \xi_{ij} = \lambda_{ij}{\sqrt{m_im_j}\over{v}}~,
\label{coup}
\end{equation} 
where $v=246$ GeV is the weak scale, 
and the $\lambda_{ij}$ are naturally all of ${\cal O}(1)$.  
This ansatz has no conflict with current phenomenological
observation, and yet predicts rich physics in the heavy-flavor
generations. Under this ansatz, the
top-charm-scalar coupling $\xi_{ct}$ is the biggest in the model, 
and is approximately $0.05$.  

Model III is fairly general.  A number of more specific models have a
large TCH coupling as well. 
Burdman \cite{burdman} studied a topcolor model \cite{hill},
which gave a large TCH coupling of the same order as Model III 
(although his ``Higgs"
boson had a mass larger than that of the top quark).  A topflavor
model \cite{heyuan} analyzed the flavor-changing coupling of the
top and
charm to a top-pion, finding a TCH coupling that is substantially larger
than that
of Model III. Recently, a bosonic topcolor model \cite{aranda} was
proposed that
automatically has the TCH coupling as the {\it largest} fermionic coupling
of the
Higgs boson.  An extensive analysis of models with additional singlet
quarks ($Q$) by Higuchi and Yamamoto \cite{higuchi} gives a
flavor-changing coupling which is proportional to the product of the $cQ$
and
$tQ$ mixings; this can be larger than the Model III coupling.  
In general, in
models which treat the top quark differently than other quarks
motivated by its weak-scale mass, one might expect TCH coupling of 
Model III size or larger.  

There have been numerous studies of Model III with the above ansatz,
examining $K,\ B,\ \mu$ and $\tau$ decays, as well as $K^0$, 
$B^0$ and $D^0$ mixing \cite{sheryuan}.  Here, however, 
we will be concerned with the large TCH coupling only.

The rate for the decay $t\rightarrow c h$ was first calculated by 
Hou \cite{hou92}. For a top-quark mass of $175$ GeV, the
branching ratio of $t\rightarrow c h$ to $t\rightarrow b W^+$ is

\begin{eqnarray}
\label{BR}
{\Gamma(t\rightarrow c h)\over \Gamma(t\rightarrow b W^+)}
&\approx&\lambda_{ct}^2\ \frac{m_c}{m_t}\ \frac{(1-{m_h^2 \over m_t^2})^2}
{(1-\frac{m_W^2}{m_t^2})\,(1-\frac{m_W^2}{m_t^2}-2\,
\frac{m_W^4}{m_t^4})} \\ \nonumber
&\approx& 0.009\ 
\lambda_{ct}^2\ \left(1-{m_h^2\over m_t^2}\right)^2.
\end{eqnarray}
This implies that the branching fraction of $t\to ch$ can be typically
of $10^{-3}$ or higher for $\lambda_{ct}\sim {\cal O}(1)$. 
If the Higgs is heavier than the top, one can have 
$h\rightarrow \bar{t}c+\bar{c}t$, as first discussed
in Ref.~\cite{hou92}. This leads to the process $e^+e^-\rightarrow
h^0A^0\rightarrow
\bar{t}\bar{t}cc+\bar{c}\bar{c}tt$ at a linear collider \cite{houlin}, 
to like-sign top-pair
production at the LHC \cite{houlinmayuan}, and to the linear collider process
$e^+e^-\rightarrow \bar{\nu}_e\nu_e(\bar{t}c+\bar{c}t)$ \cite{barshalom}. 
These processes all assume a Higgs boson heavier than the top, however,
and recent
 electroweak precision data \cite{electroweak}, as well as hints from
LEP \cite{lep},
prefer the scenario of a rather light Higgs boson, 
making $t\rightarrow c h$
kinematically accessible.   
As one can see from Eq.~(\ref{BR}), the branching ratio for
$\lambda_{ct}=1$ and a $115$ GeV Higgs mass is $3 \times 10^{-3}$.

If one looks  at one-loop processes, there are many more
possibilities.  Most extensions of the Standard Model 
will avoid the GIM suppression that makes the 
coupling so small in the Standard Model \cite{mele}.   
The most popular extension is the MSSM.  
In this model, there {\it must} be a Higgs boson lighter
than the top quark, and so the decay will occur.  An extensive analysis of 
$t\rightarrow c H$, where
$H$ refers to any one of the three neutral scalars in the MSSM, was
carried out in Ref.~\cite{yangli}.  They show that the dominant
contribution comes about from loops with gluinos and squarks, and the
branching ratio
ranges from
$10^{-5}$ to $5 \times 10^{-4}$ over the MSSM parameter space.   In an
R-parity
violating SUSY model, a one-loop contribution can also give \cite{ehyz} a
branching ratio as large as $10^{-5}$.

We see that models with a tree-level TCH coupling have branching ratios of
the order
of $10^{-3}-10^{-2}$, and models with a one-loop coupling can have
branching ratios
of the order of $10^{-5}-10^{-4}$.   
What level is experimentally detectable at high energy colliders?  
Recently, discovery
limits at the LHC were calculated \cite{aguilar} and are approximately
$5\times 10^{-5}$. While the results are quite encouraging, detailed
Monte Carlo simulation would be needed to draw a definitive conclusion
due to the complicated background issue at hadron colliders.
In a linear collider, backgrounds are much more manageable, although 
signal cross sections are lower.  
In this paper, we assume that a light Higgs boson will be 
observed and examine in detail the discovery potential for 
its FCNC coupling at a linear collider.

%%%%%%%
%figure 1
%%%%%%%

\begin{figure}[tb]
\begin{center}
\begin{picture}(240,100)(0,0)
\ArrowLine(10,20)(40,50)
\Text(10,12)[]{$e^-$}
\ArrowLine(40,50)(10,80)
\Text(10,88)[]{$e^+$}
\Photon(40,50)(80,50){4}{4}
\Text(60,62)[]{$\gamma,Z$}
\ArrowLine(110,20)(80,50)
\Text(110,12)[]{$\bar{t}$}
\ArrowLine(80,50)(110,80)
\Text(110,88)[]{$c$}
\dashline{5}(95,65)(110,50)
\Text(110,42)[]{$h$}
\ArrowLine(130,20)(160,50)
\Text(130,12)[]{$e^-$}
\ArrowLine(160,50)(130,80)
\Text(130,88)[]{$e^+$}
\Photon(160,50)(200,50){4}{4}
\Text(180,62)[]{$\gamma,Z$}
\ArrowLine(230,20)(200,50)
\Text(230,12)[]{$\bar{t}$}
\ArrowLine(200,50)(230,80)
\Text(230,88)[]{$c$}
\dashline{5}(215,35)(230,50)
\Text(230,58)[]{$h$}
\end{picture}
\end{center}
\caption{Tree level Feymann diagrams for $\epem \to \bar{t}ch$ process.
  Diagrams for $\epem \to t\bar{c}h$ process are similar.}
\label{fey}
\end{figure}
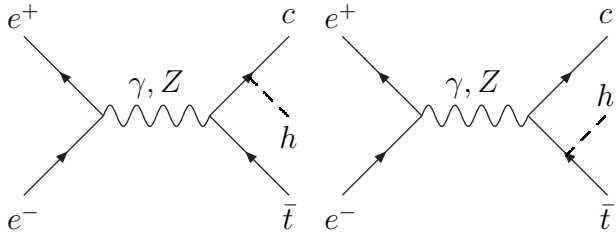

\section{Top-charm-Higgs coupling at a linear Collider}

\subsection{Production Cross Section}

\begin{figure}[tb]
\centerline{\psfig{file=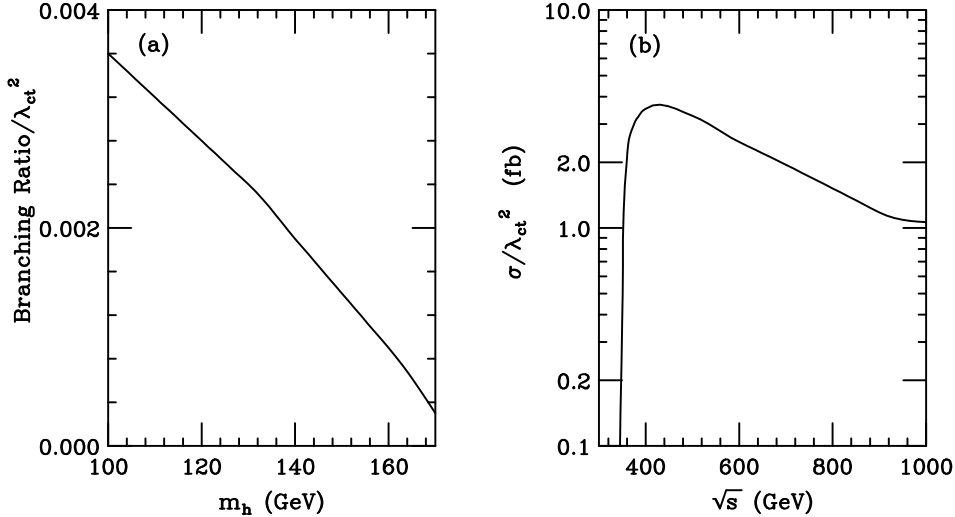,width=5.0in}}
\bigskip
\caption{(a) Branching ratio of
$t \to ch$ decay as a function of $m_h$ and (b) Cross section for 
$t\bar c h+\bar tch$ production as a function of
center-of-mass energy in $\epem$ collisions for $m_h = 120$ GeV.} 
\label{total}
\end{figure}

Using the coupling constant given in Eq.~(\ref{coup}), 
we wish to explore the limits 
obtainable on $\lambda_{ct}$ at a linear collider.
We consider the process 
\begin{equation}
\label{tbch}
\epem \to \bar{t}ch,\ t\bar{c}h~,
\end{equation}
where the corresponding Feynman diagrams are shown in Fig.~\ref{fey}.
The dominant contribution to process~(\ref{tbch})
is from $\epem \to t\bar{t}$ followed by $t\, (\bar t)$ decay into 
$ch\, (\bar ch)$, namely 
$\sigma(\epem\to t\bar{t})\times [Br(t \to ch)
+Br(\bar{t} \to \bar{c}h)]$.
The decay branching ratio is calculated and shown in
Fig.~\ref{total}(a), where we see that for $m_h = 120$ GeV and 
$\lambda_{ct} = 1$, $t \to ch$ 
has a branching ratio of $2.8\times10^{-3}$. 
At tree level, top-pair production 
has a total cross section of $580$ fb at the center-of-mass
energy $\sqrt s=500$ GeV. Thus each channel will
have a cross section of $1.6$ fb.
The total cross section of $\epem \to \bar{t}ch$ and $t\bar{c}h$
is presented in Fig.~\ref{total}(b) as a function of $\sqrt{s}$,
where and henceforth $m_h = 120$ GeV is chosen for illustration.  
The total cross section scales as $\lambda_{ct}^2$ and is of
order of 1 fb for $\lambda_{ct} = 1$.

%%%%%%%
%figure 2
%%%%%%%

\subsection{Signal And Background}

Since we are motivated to consider a light Higgs boson
with a mass around 120 GeV,  
we concentrate on the process in which the Higgs boson
decays to $b\bar{b}$ and 
$t$ decays into $bW$.  The $W$ can then undergo either
hadronic (2 jets) or leptonic ($\ell^\pm\nu$) decays. 
The $h \to b\bar{b}$ branching ratio for $m_h = 120$ GeV
is about $85\%$.
The signal thus consists of the following channels
in the final state
\begin{eqnarray}
\label{hadr}
\epem &\to& b\bar{b}, \, b + 3\,jets \\
{\rm or}\quad &\to& b\bar{b}, \, b\, \ell^\pm \nu + 1\,jets.
\label{lept}
\end{eqnarray}

The primary background to the above signal is from the SM top-pair
production $\epem \to t\bar{t} \to \bar{b}W^+ bW^-$, with one $W$
decaying hadronically and the other decaying either hadronically or 
leptonically, depending on which signal of Eqs.~(\ref{hadr}) 
and (\ref{lept}) we are looking at.  
In order to identify the signal from the background, we
require that 3 $b$'s be tagged. The efficiency for a single $b$
tagging is taken to be $65\%$ \cite{jackson}. This still does not
eliminate the background: For the case that both $W$'s decay 
hadronically, one out of the four non-$b$ jets from $W$ decay
to light quarks may be misidentified 
as a $b$-jet. We assume the misidentification to be
$1\%$ for each jet.  Thus for four jets the total misidentification
probability is $4\%$. Similarly, for the case that one of the two 
$W$'s decays leptonically, one of the two non-$b$ 
jets from $W$ decay to light quarks is misidentified as a
$b$-jet.  The SM background induced by one $W$ decay into $\bar{c}b$ 
is negligible due to the small size of the CKM mixing matrix 
element $V_{cb}\sim 0.04$.  The entries of the first row in 
Table~\ref{rates} show the signal and background 
cross sections at $\sqrt{s} = 500$ GeV before applying any 
kinematical cuts but including decay branching fractions and 
$b$-tagging efficiencies. We find that the background rate is
still overwhelming.
%%%%%%%
% Table 1
%%%%%%%
\begin{table}[tb]
\begin{center}{
\begin{tabular}{@{\hspace*{0.5cm}}ccccc@{\hspace*{0.5cm}}}
& signal&bg(hadronic)&bg(leptonic)&bg(total) \\ \hline
no cuts & 0.75 & 6.56 & 3.23 & 9.79 \\
basic cuts & 0.60 & 0.28 & 0.26 & 0.54 \\
addi. cuts & 0.52 & 0.13 & 0.12 & 0.25 \\ 
\end{tabular}
\caption[]{Cross sections in fb at $\sqrt s=500$ GeV
for signal ($m_h=120$ GeV and $\lambda_{ct} = 1$) 
and hadronic, leptonic and total backgrounds 
before and after cuts.}
\label{rates}}
\end{center}
\end{table}
%\vskip -0.5cm

%%%%%%%%%
% figure 3 
%%%%%%%%%

\begin{figure}[tb]
\centerline{
\psfig{file=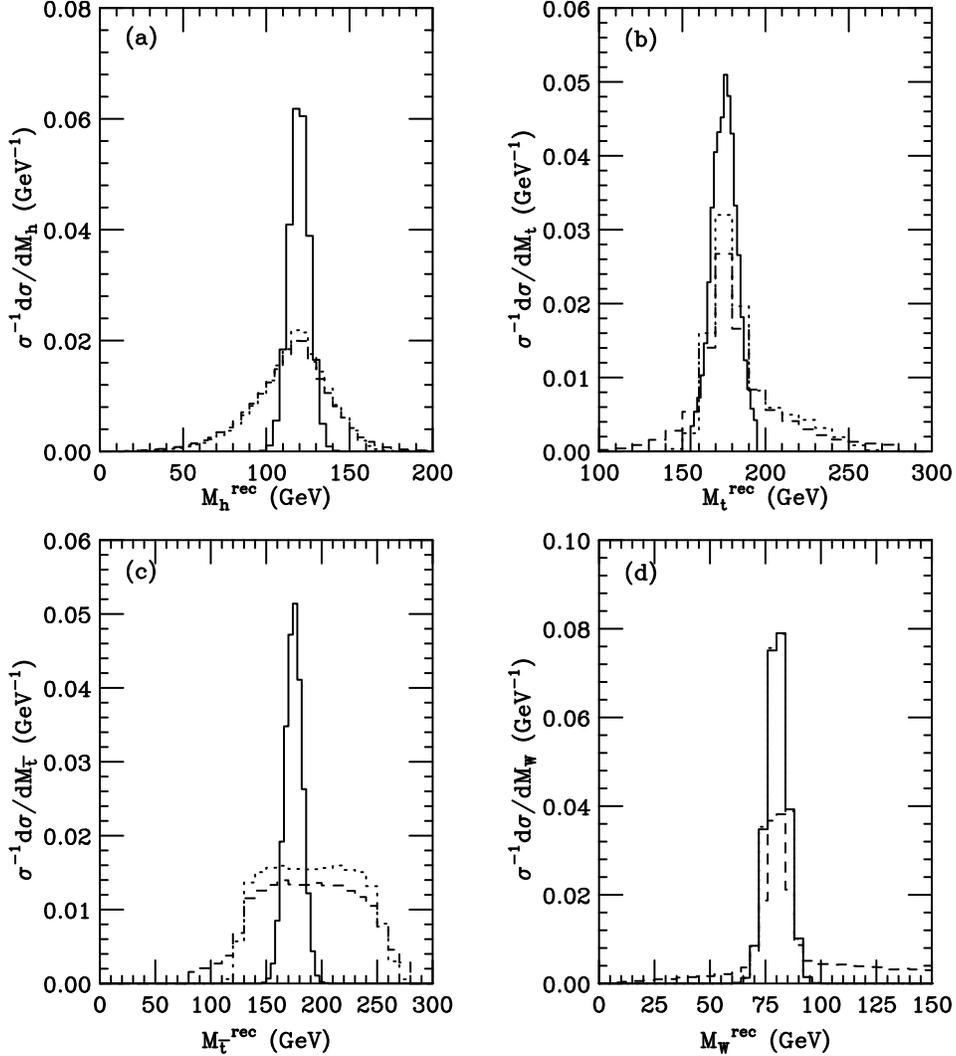,width=5in}}
\caption{Normalized distribution for the signal (solid), hadronic 
background (dashes) and leptonic background (dots) with respect to the 
reconstructed masses (a) $M_h^{rec}$, (b) $M_t^{rec}$, 
(c) $M_{\bar{t}}^{rec}$ and (d) $M_W^{rec}$,
with $\sqrt s=500$ GeV and $m_h=120$ GeV.}
\label{crucial}
\end{figure}

We next try to make use of the distinctive signal kinematics by
reconstructing $h$, $t$, $\bar{t}$ and $W$ masses from the 
final state momenta. To simulate the detector response, we smear
the jet energies according to a Gaussian spread
\begin{equation}
{\Delta E\over E}={45\%\over \sqrt E} \oplus 2\%~,
\end{equation}
where $\oplus$ denotes a sum in quadrature.
For the signal, we first consider the case 
where $W$ decays into two jets. Of the three tagged $b$'s,
we choose two $b$'s whose invariant mass is closest to $m_h$ and identify 
them as the $b$'s coming from $h$ decay ($M^{rec}_h$).  
Then from the three light jets we pick the jet that,
together with $b\bar{b}$ selected above, produces an invariant mass 
($M_t^{rec}$) that is closest to $m_t$.  
This jet is now identified as ${c}$ from 
$t\to {c} b\bar b$ decay.  The remaining two jets can be used to 
construct the $W$ mass $M^{rec}_W$ and with the third $b$, 
we have $M^{rec}_{\bar{t}}$.  
If $W$ decays into $l\nu$, it is simpler.  We simply need to choose 
two of the three $b$-jets to construct $M^{rec}_h$. The only non-$b$ 
jet is identified as ${c}$.  We get $M^{rec}_W$ from the momenta of 
$\ell^\pm$ and missing energy assigned to $\nu$, 
after adding the momentum of the third $b$, we have $M^{rec}_{\bar{t}}$. 
For the background in which $W$ decays hadronically, we randomly pick one
from the 4 non-$b$ jets and assume it to be misidentified as a $b$-jet.
Then we follow the same construction as for the signals, 
we will get $M_h^{rec}$, $M^{rec}_t$, $M^{rec}_{\bar{t}}$ and 
$M^{rec}_W$. For the background when $W$ decays 
leptonically, we randomly choose one of the two non-$b$ jets to be 
misidentified as $b$ and again follow the analysis for the signals to 
get the four reconstructed invariant masses.  The normalized differential
distribution with respect to these reconstructed masses are shown in
Fig.~\ref{crucial} for signals and backgrounds.  The characteristic
difference between the signal and background is evident from
these mass distributions. The first crucial reconstructed mass is 
$M^{rec}_h$. A peak structure in this variable provides 
the confidence of the signal
observation and gives the discrimination power between the signal 
and background. The wide width is due to the jet-energy smearing
discussed earlier.
The other distinctive mass variable is $M^{rec}_t=M_{bbc}$,
which reconstructs $m_t$ for the signal and no structure for the
background due to the incorrect jet combination. Similarly, 
$M^{rec}_{\bar t}\approx m_t$ for the signal but it spreads out
for the background, as seen in Fig.~\ref{crucial}(c). 
Those distributions
motivate us to device more judicial kinematical cuts.
Our basic kinematical cuts are listed in Table~\ref{cuts} 
based on these four reconstructed masses.

%%%%%%%%
% table 2
%%%%%%%%
\begin{table}[tb]
\begin{center}
\begin{minipage}{4in}
\begin{center}{
\begin{tabular}{@{\hspace*{1.5cm}}cc@{\hspace*{1.5cm}}}
%\begin{tabular}{cc}
basic cuts & additional cuts \\ \hline
$110 < M_h^{rec} < 130$ & $20 < E_c^{lab} < 130$ \\
$160 < M_t^{rec},\ M_{\bar{t}}^{rec} < 190$ & $40 < E_c^{rest}<60$\\
$65  < M_W^{rec} < 95$ & $E_b^{min} < 120,\ 80 < E_b^{max}$ \\
\end{tabular}
\caption{The basic and additional kinematical cuts applied
to our signal-to-background optimization (all values in GeV).}
\label{cuts}}
\end{center}
\end{minipage}
\end{center}
\end{table}

%\vskip -0.5cm

\begin{figure}[tb]
\centerline{\psfig{file=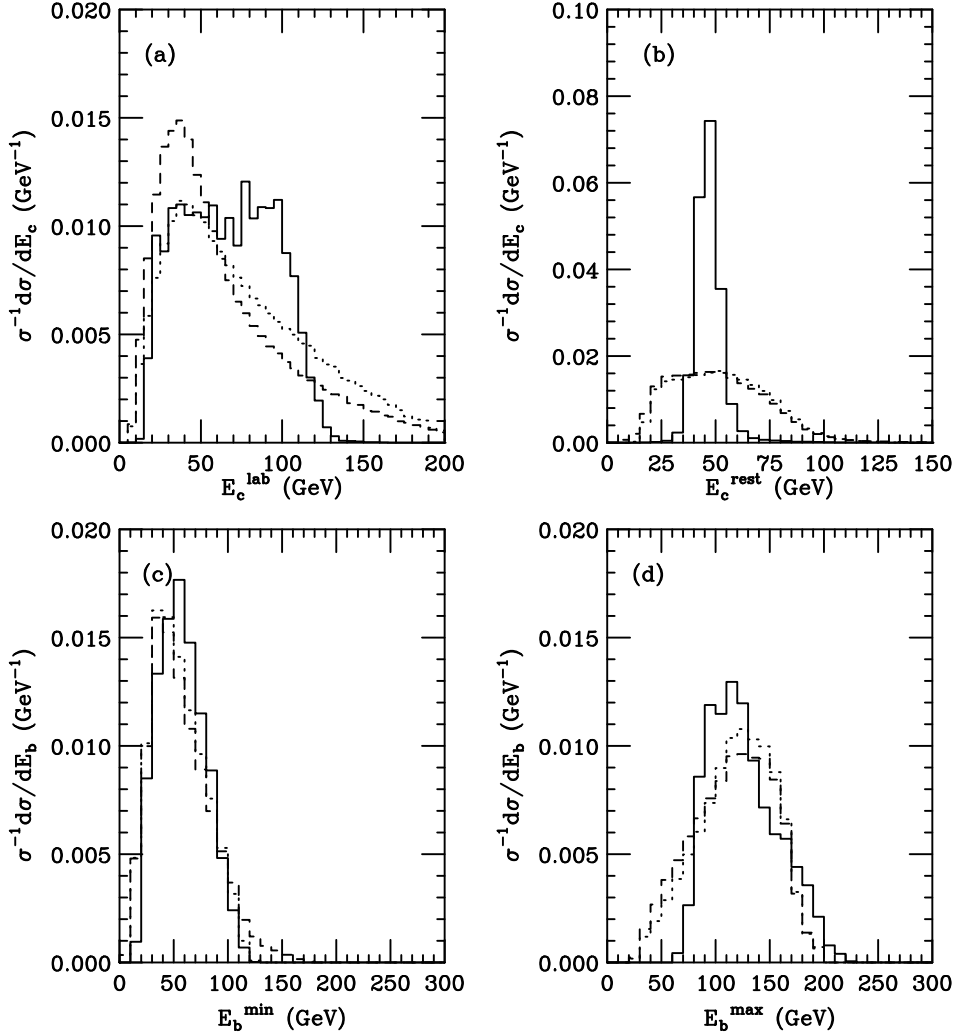,width=5in}}
\caption{Normalized distributions for signal (solid), hadronic 
background (dashes) and leptonic backgrounds (dots) with respect to
(a) $E_c^{lab}$, (b) $E_c^{rest}$, (c) $E_b^{min}$
and (d) $E_b^{max}$, with $\sqrt s=500$ GeV and $m_h=120$ GeV.}
\label{addi}
\end{figure}

Additional cuts can be applied to further 
increase the signal-to-background ratio. We first notice that
the charm-jet energy from the top decay is monotonic in the top-rest 
frame as a result of a two-body decay
\begin{equation}
E_c^{rest}={m_t\over 2}(1-{m_h^2\over m_t^2}),
\label{ec}
\end{equation}
which is about 50 GeV for $m_h=120$ GeV. In contrast, 
the faked charm jet from $W$ decay will 
have an energy more spread out. 
The normalized distributions are shown in Fig.~\ref{addi}.
Figure \ref{addi}(a) is
the $E_c$ distribution in the $e^+e^--$lab frame which presents
a nearly clear range due to the Lorentz boost of the top quark
motion, where the sharp end-point in $E_c$ is sensitive
to $m_h$, that may provide an independent kinematical
determination for $m_h$. 
Figure \ref{addi}(b) shows the $E_c$ distribution
in the top-quark rest frame based on the $t\to b\bar bc$
reconstruction. We see that $E_c$ is monotonic near 50 GeV
modulus to the jet-energy resolution. This provides a good
discriminator for the signal and background. In fact,
the reconstructed Higgs boson energy in the top-quark
rest frame should be also monotonic $E_h^{rest}={m_t\over 2}
(1+m_h^2/m_t^2)$, but it is correlated with $E_c^{rest}$. 
We also look at the two $b$-jets coming from $h$ 
decay and examine the harder and softer
ones of the two in energy, separately, as illustrated
in Figs.~\ref{addi}(c) and \ref{addi}(d).
The optimal cuts are given in Table~\ref{cuts}.  
After applying the basic and additional cuts
the signal-to-background ratio improves significantly,
as summarized in the last two rows of Table~\ref{rates}.

%%%%%%%
%Figure 4
%%%%%%%

\subsection{Sensitivity to the $tch$ coupling}
\begin{figure}[tb]
\centerline{\psfig{file=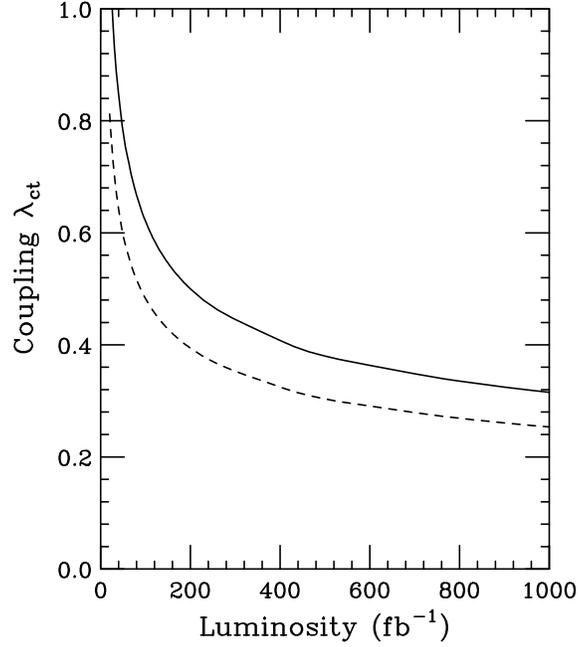,width=3in}}
\caption{3$\sigma$ sensitivity (solid) to the FCNC $tch$ couplings at 
$\sqrt s=500$ GeV for $m_h=120$ GeV as a function of integrated 
luminosity.  The dashed curve is for $2\sigma$ sensitivity.}  
\label{sensi}
\end{figure}

Given the efficient signal identification and substantial background 
suppression achieved in the previous section, 
we now estimate the sensitivity to the FCNC $tch$
coupling from this reaction using Gaussian statistics, which is applicable
for large signal event samples. We define the statistical significance by
\begin{equation}
\sigma = {N_S\over \sqrt{N_S+N_B}},
\end{equation}
with $N_S$ and $N_B$ being the number of signal and background
events. We note that $\sigma=3$ (called 3$\sigma$) approximately 
corresponds to a $99\%$ confidence level.
Figure~\ref{sensi} presents the 3$\sigma$ (2$\sigma$) sensitivity to the FCNC 
couplings by the solid (dashed) curve as a function of integrated luminosity
for $\sqrt s=500$ GeV and $m_h=120$ GeV. Recall that at 
$\lambda_{ct} =1$, the $t \to ch$ branching ratio is 
about $2.8\times10^{-3}$. Such an order of magnitude can be
anticipated in models with tree-level FCNC couplings, and
can be easily observed at a LC with an integrated luminosity
of less than 40 fb$^{-1}$. As $\lambda_{ct}$ varies, 
the branching ratio scales like $2.8\times10^{-3} \lambda_{ct}^2$.
For 1-loop induced FCNC decays such as in SUSY models, the 
branching ratios can be about $10^{-5} - 5\times 10^{-4}$, 
corresponding to $\lambda_{ct}$ of $0.06-0.4$. A linear collider
with $500\ {\rm fb}^{-1}$ integrated luminosity
will begin to be sensitive to this range of the coupling 
of $\lambda_{ct}\approx 0.4\ (0.3)$ at a 3$\sigma$ (2$\sigma$)
level, but a higher luminosity will be needed to extend the 
coverage of the parameter space to a level about 0.2.

%%%%%%%
% Figure 5
%%%%%%%

\section{Conclusions}

Models with a top-charm-Higgs coupling come in two categories: Those with 
a tree-level coupling and those with a coupling induced at one-loop.  In the 
former case, such as the Model III two-Higgs doublet model, one expects 
$\lambda_{ct}$ to be of ${\cal O}(1)$.  
If the Higgs is somewhat lighter than the top quark,
then the $t\rightarrow ch$ decay will be detected within the first 
$40\ {\rm fb}^{-1}$ of running of a linear collider, 
as indicated in Fig.~\ref{sensi}.
In the latter case, such as supersymmetric models (in which we {\it know} 
that one of the Higgs bosons will be sufficiently light), 
the branching ratios can be in the range of 
$10^{-5}$ to a few times $10^{-4}$.
This corresponds to a $\lambda_{ct}$ of $0.06-0.4$. A linear collider
with $500\ {\rm fb}^{-1}$ integrated luminosity
will begin to be sensitive to this range of 
the coupling at a 3$\sigma$ level, 
and higher luminosity will be needed to substantially extend the 
coverage of the parameter space.

\vskip 0.2cm
{\it Acknowledgments}:
The work of T.H. and J.J. was supported in part by
a DOE grant No.~DE-FG02-95ER40896 and in part by 
the Wisconsin Alumni Research Foundation.  The work of M.S. was supported
in part by an NSF grant No.~PHY-9900657.

\end{document}